\documentclass{aastex631}
\usepackage{bm}%! To use \bm{} command (bold italic for the vector)
\usepackage{color}

\newcommand{\msun}{M_\odot}

\newcommand{\cc}{{\rm cm^{-3}}}
\newcommand{\dg}{^\circ}

\shorttitle{Twisted Pseudodisk}
\shortauthors{Machida, Hirano and Basu}
\begin{document}
\title{Twisted Pseudodisk and Asymmetric Mass Accretion on the Circumstellar Disk}

%%\correspondingauthor{Masahiro N. Machida}
%%\email{machida.masahiro.018@m.kyushu-u.ac.jp}

\author[0000-0002-0963-0872]{Masahiro N. Machida}
%%\author{Masahiro N. Machida}
\affiliation{Department of Earth and Planetary Sciences, Faculty of Science, Kyushu University, Fukuoka 819-0395, Japan}
\affiliation{Department of Physics and Astronomy, University of Western Ontario, London, ON N6A 3K7, Canada}
\email{machida.masahiro.018@m.kyushu-u.ac.jp}

\author[0000-0003-0855-350X]{Shantanu Basu}
%%\author{Shantanu Basu}
\affiliation{Canadian Institute for Theoretical Astrophysics, University of Toronto, 60 St. George St., Toronto, ON M5S 3H8, Canada}
\affiliation{Department of Physics and Astronomy, University of Western Ontario, London, ON N6A 3K7, Canada}
\email{basu@uwo.ca}

\author[0000-0002-4317-767X]{Shingo Hirano}
%%\author{Shingo Hirano}
\affiliation{Department of Applied Physics, Faculty of Engineering, Kanagawa University, Kanagawa 221-0802, Japan}
\email{hirano0613@gmail.com}

\begin{abstract}
We model gas inflow patterns onto circumstellar disks and the evolution of the pseudodisk using three-dimensional resistive MHD simulations. 
Starting from a prestellar core without turbulence and with a misalignment between the initial magnetic field and rotation axis, the simulations are performed for  $\sim10^5$\,yr after protostar formation.
After disk formation, the magnetic field around the disk becomes significantly distorted due to the disk rotational motion.
Consequently, the structure of the pseudodisk also evolves into a complex morphology.
As a result, both accretion onto the disk and outflow become asymmetric and anisotropic.
Accretion to the disk occurs primarily through narrow-channel flows or streams.
The time evolution of the infalling envelope leads to non-steady accretion onto the disk, which in turn causes variability in the mass accretion onto the central protostar.
This study demonstrates that complex infalling envelope structures and channelized accretion flows onto the disk naturally arise even without assuming turbulence or external asymmetric inflows.
%%In idealized cases where the magnetic field and rotation axis are perfectly aligned, gas accretion proceeds in an axisymmetric and isotropic manner. However, such alignment is unlikely under realistic conditions.
\end{abstract}

\keywords{
Magnetohydrodynamical simulations (1966) ---
Protostars (1302) ---
Protoplanetary disks (1300) ---
Circumstellar disks (235) ---
Stellar jets (1807) ---
Star formation (1569) 
}

%%%%%%%%%%%%%%%%%%%%%%%%%%%%%%%%%%%%%%%%
\section{Introduction}
\label{sec:intro}
%%Magnetic fields play a crucial role in the process of star formation.
%%They drive jets and outflows, which remove excess angular momentum from the central region and transport it into the interstellar medium \citep{Pudritz1986, Matzner2000, machida13}.
%%Dissipation of magnetic fields near the protostar weakens magnetic braking, enabling the formation of a circumstellar disk \citep{tsukamoto23b}.

Observations have revealed that molecular cloud cores, which are the birthplaces of stars, are threaded by coherent magnetic fields \citep[e.g.,][]{Planck2016}.
In such cores, the magnetic energy dominates over the rotational energy and is comparable to the gravitational energy \citep{crutcher10,Caselli2002}. 
As a result, gravitational collapse preferentially occurs along magnetic field lines, which leads to the formation of a flattened, disk-like structure known as a pseudodisk \citep{Galli1993}.
The pseudodisk persists even after the formation of a protostar and surrounds the circumstellar disk as an extended structure \citep[e.g.,][]{Basu24}.
Although the pseudodisk forms due to the anisotropic Lorentz force, it is not supported against gravity and therefore continues to collapse.
The contraction proceeds more slowly than in the non-magnetized case due to the influence of the magnetic field \citep{Hirano2025}.  
Accretion onto the disk or protostar occurs through the pseudodisk, which corresponds to the infalling envelope \citep{machida13}.

Recent high-resolution ALMA observations have revealed complex structures around circumstellar disks \citep{tokuda14,tokuda16,Yen2019,Pineda2020}.
Some of these features can be explained by interchange instability occurring at the outer edge of the disk \citep{tokuda23,Tokuda24,fielder24,tanious24}, but more extended and irregular structures have also been detected, which cannot be accounted for by this mechanism alone \citep[for details, see][]{pineda23}.
Moreover, asymmetric gas accretion onto circumstellar disks has been reported \citep[e.g.,][]{Garufi2022,Valdivia-Mena2022,Choudhury2025,Kido025}. 
In the conventional picture, accretion onto the disk via the pseudodisk is expected to occur in an axisymmetric %or spherically symmetric
fashion \citep{Tomisaka1995,Tomisaka1996,Basu1994,Basu1995,Basu1995b,machida05a,machida06}.
Consequently, recent observations of asymmetric accretion have often been interpreted as resulting from gas inflow outside the molecular cloud core, the infall of small gas clumps, or internal turbulence within the core \citep{pineda23}.

Many previous simulation studies about collapsing star-forming cores have assumed, for simplicity, that the angular momentum vector of the molecular cloud core is aligned with the global magnetic field threading the core \citep[see review by][]{tsukamoto23b}.
However, both observations and simulations suggest that these directions are not necessarily parallel \citep[e.g.,][]{Hull2013,Misugi2024}.
We have previously studied the early evolution of molecular cloud cores with misaligned magnetic fields and angular momentum vectors using three-dimensional simulations, focusing on disk formation and the development of outflows and jets shortly after protostar formation \citep{Kataoka2012,Shinnaga2012,Hirano2019,hirano20,Machida2020d}. 
However, the long-term evolution of the pseudodisk and the pattern of accretion onto the disk have not been explored in detail. In this study, we investigate the long-term evolution of pseudodisks and accretion patterns in collapsing molecular cloud cores, where the global magnetic field is initially misaligned with the angular momentum vector.

\section{Numerical settings and Initial Conditions}
Since the numerical settings and initial conditions are almost the same as our previous studies \citep{Kataoka2012,Hirano2019,Machida2020d}, we describe them briefly.  
As the initial condition, we adopt a critical Bonnor–Ebert sphere with a central density $n_{c,0} = 10^4\,\cc$ and an isothermal temperature $T = 10\,\mathrm{K}$.
To induce gravitational collapse, the density is enhanced by a factor of $f = 1.68$ \citep{machida24}.
The mass and radius of the initial cloud are $M_{\rm cl,0} = 8.1\,\msun$ and $R_{\rm cl,0} = 4.8 \times 10^4\,\mathrm{au}$, respectively.
The computational boundary is placed at a distance $16\,R_{\rm cl,0}$ from the cloud center to suppress reflections of Alfvén waves generated within the cloud ($r < R_{\rm cl,0}$) \citep{machida13}.
Gravity is turned off outside the initial cloud ($r > R_{\rm cl,0}$), and inflow from the ambient medium is prohibited \citep{machida20}.
In contrast, the mass outflow from the cloud ($r < R_{\rm cl,0}$) to the interstellar region ($r > R_{\rm cl,0}$) is allowed to adequately account for protostellar outflows \citep{Basu24}.

A uniform magnetic field $B_0 = 5.5 \times 10^{-6}\,\mathrm{G}$ is imposed across the entire computational domain.
The mass-to-flux ratio of the initial cloud is $\mu_0 = 3$, normalized by the critical value $(2 \pi G^{1/2})^{-1}$.
Rigid rotation with an angular velocity $\Omega_0 = 2.5 \times 10^{-14}\,\mathrm{s}^{-1}$ is imposed within the initial cloud ($r < R_{\rm cl,0}$).
The ratios of thermal, rotational, and magnetic energies to the gravitational energy are $\alpha_0 = 0.5$, $\beta_0 = 0.02$, and $\gamma_0 = 0.1$, respectively. 
Each energy component is numerically calculated within the initial cloud ($r < r_{\rm cl,0}$).
%The angle $\theta_0$ between the initial magnetic field and angular momentum vectors is treated as a free parameter.
In this paper, we present the results for the model with an angle $\theta_0 = 30\dg$ between the initial magnetic field direction and angular momentum vector.
%RThe results for models with other values of $\theta_0$ will be presented in a companion paper.

To follow the cloud evolution, we use our nested-grid code \citep{machida04,machida05a,machida10,machida14,machida21}.
Each cubic grid consists of $(i, j, k) = (64, 64, 64)$ cells.
Initially, five grid levels ($l = 1$–$5$) are nested, where the level of refinement is denoted by the index $l$.
The initial cloud is embedded in the fifth grid level ($l = 5$), which has a box size of $L(5) = 9.5 \times 10^4\,\mathrm{au}$ and a cell width of $h(5) = 1.5 \times 10^3\,\mathrm{au}$.
The maximum grid level is set to $l = 16$.
The coarsest grid ($l = 1$) has a box size of $L(1) = 1.5 \times 10^6\,\mathrm{au}$ and a cell width of $h(1) = 2.3 \times 10^4\,\mathrm{au}$.
The finest grid ($l = 16$) has a box size of $L(16) = 46\,\mathrm{au}$ and a cell width of $h(16) = 0.73\,\mathrm{au}$.
As the cloud begins to collapse, finer grids are automatically generated to ensure that the local Jeans wavelength is resolved by at least 16 cells.

When the density exceeds $n_{\rm sink} = 10^{13}\,\cc$ in the collapsing cloud, a sink cell is introduced  at the center of the computational domain, and its position is kept fixed throughout the calculation \citep{machida10}. 
After the sink is created, 1\,\% of the gas exceeding $n_{\rm sink}$ within the sink radius of $r_{\rm sink} = 1\,\mathrm{au}$ is removed from the computational domain and added to the protostellar mass \citep{machida14,machida16}. 
The magnetic flux associated with the gas accreted onto the sink is not removed from the computational domain in order to maintain the divergence-free condition of the magnetic field. 
Because the initial conditions maintain point symmetry about the center of the computational domain, fixing the sink at the origin is justified in this study. 
However, small center-of-mass drifts may still arise from numerical round-off, and therefore this approach is not universally applicable \citep{hirano20}. 
%%We follow the evolution of the cloud for $\gtrsim 10^5\,\mathrm{yr}$ after the formation of the protostar.

%%%%%%%%%%%%%%%%%%%%%%%%%%%%%%%%%%%%%%%%
\section{Results}
\label{sec:results}

%%%%%%
% Fig. 1
%%%%%%
\begin{figure*}
\begin{center}
\includegraphics[width=1\columnwidth]{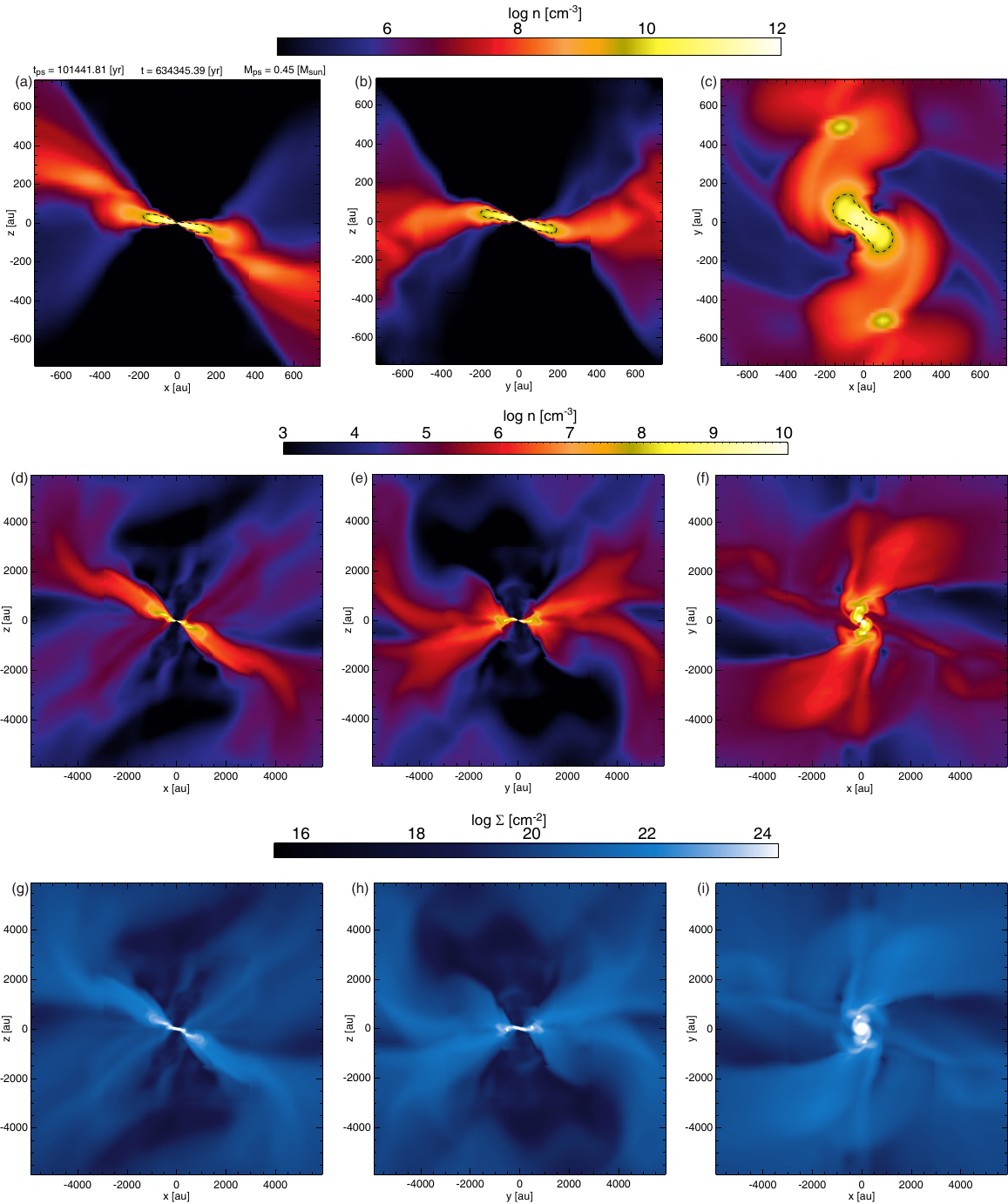}
\end{center}
\caption{
(a)–(f): Density (color) distribution on the $x = 0$ (panels (a) and (d)), $y = 0$ (panels (b) and (e)), and $z = 0$ (panels (c) and (f)) planes. 
(g)–(i): Surface density (color) distributions on the $x = 0$ (panel (g)), $y = 0$ (panel (h)), and $z = 0$ (panel (i)) planes.
The dotted curves in panels (a)–(c) indicate the rotationally supported disk.
The elapsed time $t_{\rm ps}$ after protostar formation, the time $t$ since the onset of cloud collapse, and the protostellar mass are indicated in panel (a). 
The spatial scale in panels (a)–(c) differs from that in panels (d)–(f) and (g)–(i).
An animated version of this figure is available. In the animation, the time sequence of the density and velocity distribution on the $x=0$, $y=0$, and $z=0$ plane as in Figure 1 from the beginning until the end of the simulation is shown. The duration of the
animation is 25 s. 
}
\label{fig:1}
\end{figure*}
We performed simulations of molecular cloud core evolution for approximately 100,000 years after protostar formation. %in which the initial relative angle between the magnetic field and the rotation axis was set to $\theta_0 = 30\,\dg$.  
During this period, the protostar increased its mass to $0.45\,\msun$.
Figure~\ref{fig:1} shows the density distribution at the end of the simulation.  
The top and middle panels display cross sections on the $y=0$, $x=0$, and $z=0$ planes, centered on the protostar and disk region, at spatial scales of $\sim1,000$\,au (top) and $\sim10,000$\,au (middle).  
The bottom panels show the corresponding surface density distributions.
A rotationally supported disk is indicated by the dotted curves in panels (a)–(c)
\footnote{
The rotationally supported disk is identified through the following procedure \citep{Tomida17,hirano20}.
First, we calculate the angular momentum vector of the high-density region ($n > 10^8\,{\rm cm^{-3}}$) and determine its direction. 
Next, on the plane perpendicular to this angular momentum vector, we evaluate the rotational velocity of the gas. 
We define the disk density threshold $n_{\rm disk}$ as the density above which the rotational velocity exceeds 80\% of the local Keplerian velocity. 
Finally, the disk is defined as the region where the density is greater than $n_{\rm disk}$.
}.

As seen in Figures~\ref{fig:1}(a) and \ref{fig:1}(c), the rotationally supported disk has a size of $\sim 300$\,au, while the surrounding disk-like structure extends well beyond it. 
Figure~\ref{fig:1}(a) and \ref{fig:1}(b) show that the outer disk-like structure is misaligned with respect to the inner disk.  
The inner structure corresponds to the rotationally supported disk, while the outer structure corresponds to the pseudodisk.  
Since the rotation axis of the circumstellar disk and the normal vector of the pseudodisk are not aligned, the pseudodisk becomes twisted and deformed as the disk grows around the protostar \citep[for detailed explanation see][]{Hirano2019}.  
The clumps located above and below the center in Figure~\ref{fig:1}(c) are not gravitationally bound, and they represent transient high-density structures formed as a result of the twisting of the pseudodisk \citep{Machida2020d}.

Figures~\ref{fig:1}(d) and \ref{fig:1}(e) show that the pseudodisk (or infalling envelope) extends diagonally from the upper left to the lower right (panel d), or from the lower left to the upper right (panel e), with respect to the protostar.  
%%Figures~\ref{fig:1}(c) and \ref{fig:1}(g), which provide top-down views of the pseudodisk, indicate that the structure is not a circular or ring-like disk, but rather bar-shaped.  
The structures seen in the cross-sectional views are also visible in the surface density maps, indicating that the infalling envelope is neither axisymmetric nor spherically symmetric.  
The vertical cavities above and below the disk in Figures~\ref{fig:1}(d), \ref{fig:1}(e), \ref{fig:1}(g) \ref{fig:1}(h) are carved out by the outflow.  
It should be noted that, because only an initial misalignment between the rotation axis and the magnetic field was introduced, the overall structure remains approximately point-symmetric with respect to the protostar.

%%%%%%
% Fig. 2
%%%%%%
\begin{figure*}
\begin{center}
\includegraphics[width=1.0\columnwidth]{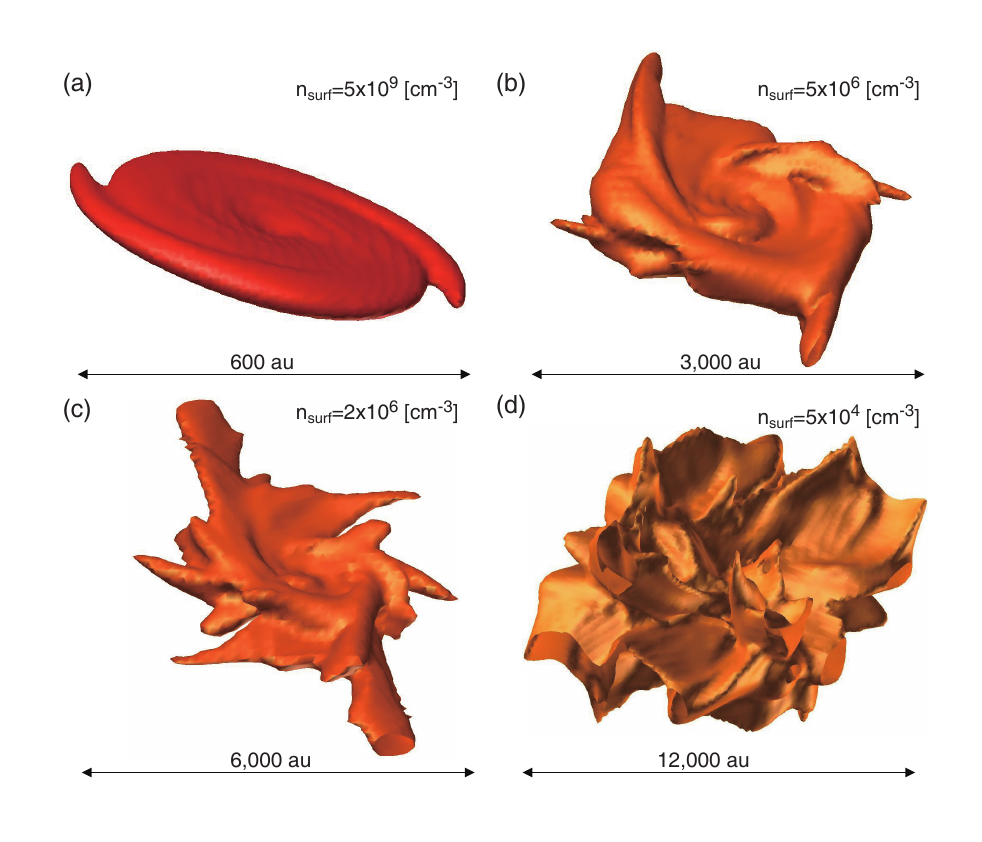}
\end{center}
\caption{
Three-dimensional view of the iso-density surface at number density $n_{\rm surf}$, as indicated in each panel, at the same epoch as in Fig.~\ref{fig:1}.
The spatial scale, also noted in each panel, differs among the panels.
}
\label{fig:2}
\end{figure*}

Figure~\ref{fig:2} shows the three-dimensional structures of the pseudodisk (infalling envelope) and the rotationally supported disk at different spatial scales.  
Each panel presents an iso-density surface at a specified number density $n_{\rm surf}$.
Figure~\ref{fig:2}(a) shows that a smooth disk structure exists within approximately 300\,au from the center, corresponding to the rotationally supported disk.  
At a larger scale of 3,000\,au, shown in Figure~\ref{fig:2}(b), the structure exhibits an overall disk-like morphology but is geometrically distorted.  
This structure is not supported by rotation and continues to collapse, with gas accreting onto the central region through narrow channels.
At the 6,000\,au scale (Fig.~\ref{fig:2}(c)), several spiky structures can be identified. These features suggest that dense gas is streaming toward the center through multiple high-density channels.  
In contrast, the regions along the disk normal direction (upper right and lower left) are less dense, indicating that gas accretion does not occur there due to the presence of outflows.
As shown in Figure~\ref{fig:1}(d), complex structures extend over a spatial range greater than 10,000\,au.  
The upward-pointing spike-shaped features in this panel correspond to the cavities carved out by the outflow.
Although the infalling envelope as a whole forms a flattened, sheet-like structure, it also exhibits complex features on smaller, localized scales \citep[see also][]{Tu2024}.

\begin{figure*}
\begin{center}
\includegraphics[width=1.0\columnwidth]{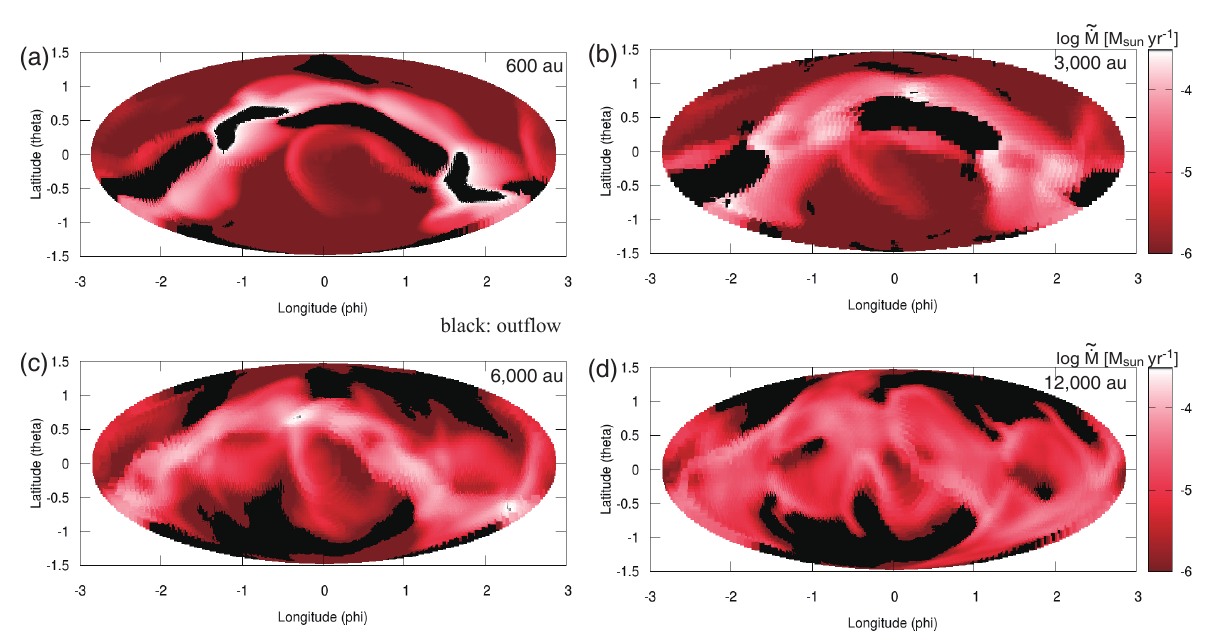}
\end{center}
\caption{
Mass inflow rates $\tilde{\dot{M}}(\theta,\phi)$ onto spherical surfaces at radii of 600\,au (panel (a)), 3,000\,au (panel (b)), 6,000\,au (panel (c)), and 12,000\,au (panel (d)), at the same epoch as in Fig.~\ref{fig:1}. 
The black regions correspond to outflowing gas with $v_r > 0$. 
The vertical and horizontal axes represent latitude 
%($-\pi < \theta < \pi$) and longitude ($-2\pi < \phi < 2\pi$), 
($-\pi/2 < \theta < \pi/2$) and longitude ($-\pi < \phi < \pi$),
respectively, in radians. 
Note that the axes correspond to spherical angles, not to the disk plane. Since the disk orientation varies with radius as the inflow moves inward, a single well-defined disk midplane cannot be drawn in this projection.
}
\label{fig:3}
\end{figure*}

Figure~\ref{fig:3} shows the mass inflow rate onto spherical surfaces centered on the protostar at different spatial scales.
In each panel, red regions indicate inflow regions with $v_r < 0$ (i.e., accretion), while black regions indicate outflow regions with $v_r > 0$.
To estimate the inflow rate, we first compute the local mass flux per unit area at each point on the spherical surface as
\begin{equation}
\dot{m}(\theta,\phi) \equiv \rho(\theta,\phi)\, v_r(\theta,\phi).
\end{equation}
Because the actual surface area associated with each sampling point varies depending on the spatial resolution of the spherical shell, we scale this local quantity to a value that would correspond to a uniform distribution 
over the entire sphere, and also present it in a clearer and 
resolution-independent form:
\begin{equation}
\tilde{\dot{M}}(\theta,\phi) \equiv \dot{m}(\theta,\phi)\, 4\pi r_0^2,
\end{equation}
where \( r_0 \) (= 600, 3,000, 6,000, and 12,000 au) is the radius of the spherical surface.
This scaled quantity should be interpreted as a diagnostic of the angular variation of the mass flux, not as the physical accretion rate.
Thus, each point in Figure~\ref{fig:3} represents the mass inflow (or outflow) rate that would be obtained if the locally measured value at that position were uniformly distributed over the entire spherical surface. 

%%The (local) inflow and outflow rates, \( \dot{M}_{\rm local} \), were calculated using the radial velocity \( v_r \) normal to a small surface area element \( dS \) and the density \( \rho \) at each location on the celestial sphere as
%%\begin{equation}
%%\dot{M}_{\text{local}} = \rho v_r dS.
%%\end{equation}
%%Subsequently, to obtain a value independent of \( dS \) for clarity, we scaled it as
%%\begin{equation}
%%\dot{M} = \dot{M}_{\text{local}} \frac{4\pi r_0^2}{dS},
%%\end{equation}
%%where \( r_0 \) (= 600, 3,000, 6,000, and 12,000\,au) is the radius of the sphere on which the mass inflow and outflow rates are calculated.
%%Therefore, each data point in Figure~\ref{fig:3} represents the mass accretion rate extended over the entire spherical surface \( 4\pi r_0^2 \) based on the local value at that point.

As seen in Figure~\ref{fig:3}, the inflow  and outflow patterns are highly complex.
However, at large scales of 12,000\,au (Fig.~\ref{fig:3}(d)) and 6,000\,au (Fig.~\ref{fig:3}(c)), gas predominantly escapes along the polar directions.
This large-scale bipolar outflow reflects the initial magnetic field configuration, which is aligned along the north-south (vertical) direction, causing the outflow to roughly follow the initial magnetic field lines.
In contrast, on smaller scales, the outflow emerges roughly along the rotation axis of the circumstellar disk and propagates in that direction \citep{Hirano2019,Machida2020d}.
As a result, as shown in Figures~\ref{fig:3}(a) and \ref{fig:3}(b), outflow regions appear as band-like structures distributed near the equatorial plane \citep[see also][]{matsumoto17}.
The inflow  rate is enhanced in regions surrounding these outflows, indicating that both inflow and outflow propagate, to some extent, along the local magnetic field lines.
Because the magnetic field lines are strongly twisted, the directions of inflow and outflow vary with spatial scale and can emerge from different regions on the sphere \citep{Machida2020d}.

%%%%%%
% Fig. 3
%%%%%%

%%%%%%
% Fig. 4
%%%%%%
\begin{figure*}
\begin{center}
\includegraphics[width=1.0\columnwidth]{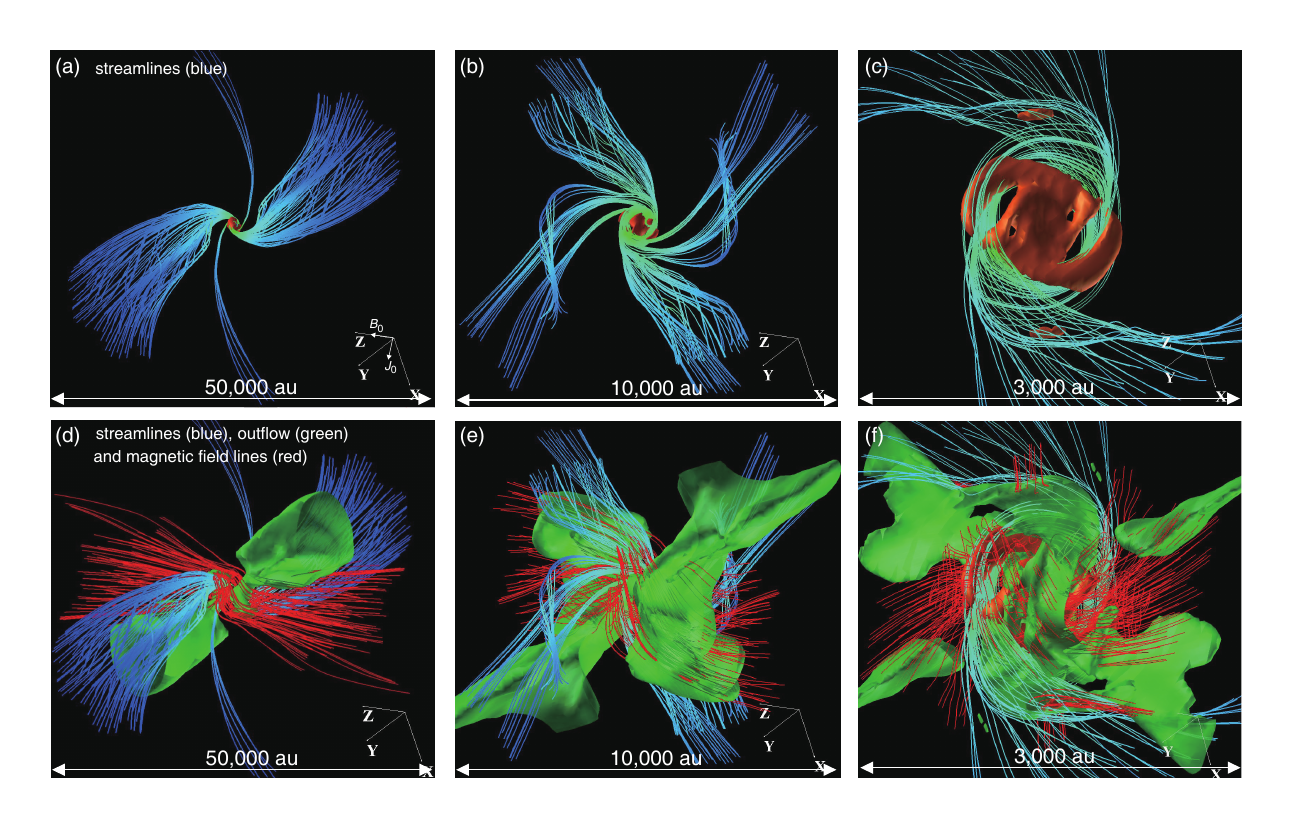}
\end{center}
\caption{
(a)–(c): Three-dimensional view of streamlines (blue lines) and high-density regions (brown), shown at different spatial scales, at the same epoch as in Fig.~\ref{fig:1}.
(d)–(f): Outflowing gas with $v_r > c_s$ (green surfaces) and  magnetic field lines (red lines) are added to the structures shown in panels (a)–(c).
%%(g)–(i): Magnetic field lines (red lines) are added to the structures shown in panels (d)–(f).
The spatial scale is indicated in each panel.
The Cartesian $x$-, $y$-, and $z$-axes are shown at the bottom right corner of each panel.
The initial directions of the magnetic field $B_0$ and angular momentum $J_0$ are indicated in panel (a). 
}
\label{fig:4}
\end{figure*}

Figure~\ref{fig:4} presents the three-dimensional patterns of infall and outflow at different spatial scales.  
In each panel, streamlines are plotted following the procedure:
\begin{itemize}
\item For each grid, the mass inflow and outflow rates are calculated on all six boundary surfaces.
\item The maximum and minimum mass infall rates, $\dot{M}_{\rm inf,\,max}$ and $\dot{M}_{\rm inf,\,min}$, are identified among all inflow cells. 
\item A threshold value $\dot{M}_{\rm inf,\,thr}$ is defined as
\begin{equation}
\dot{M}_{\rm inf,\,thr} = 10^p, \quad \text{where} \quad p = \frac{1}{2} \left[ \log_{10} \left( \dot{M}_{\rm inf,\,max} \right) + \log_{10} \left( \dot{M}_{\rm inf,\,min} \right) \right].
\end{equation}
\item Cells with infall rates exceeding the threshold ($\dot{M}_{\rm inf} > \dot{M}_{\rm inf,\,thr}$) are selected.
\item Streamlines are integrated only from these selected cells on the six boundary surfaces.
\end{itemize}
A more detailed description of how the streamlines are plotted is provided in Appendix~\S\ref{sec:A1}.
In the second row of Figure~\ref{fig:4}, regions with radial velocities exceeding the sound speed ($v_r > c_{\rm s}$) are shown as outflows (green surfaces). 
Almost all regions outside the outflow correspond to the infalling envelope, where gas is falling toward the center.
As a result, the streamlines in Figure~\ref{fig:4} highlight only those regions with relatively high infall rates within the overall infall.
As shown in Figure~\ref{fig:4}(a), the infall  pattern is highly asymmetric even at the 50,000\,au scale.  
Prominent streams directed toward the center are clearly seen from the lower left and upper right directions.  
In addition, a narrow, band-like stream runs from top to bottom.
A figure identical to Figure~\ref{fig:4}, but viewed from different angles, is provided in Appendix~\S\ref{sec:A2}.

In Figure~\ref{fig:4}(b), multiple converging streams are observed, indicating that inflow or accretion occurs primarily through several distinct channels.  
In Figure~\ref{fig:4}(c), although a disk-like structure is present at the center, the infalling  streams approach from above and below the disk plane, wrapping around its outer edges.  
In other words, inflow onto the disk occurs preferentially from specific regions near the disk midplane.

Magnetic field lines (red) are also plotted in panels (d)–(f) of Figure~\ref{fig:4}.
The field lines are integrated from the high-density regions near the center and trace the stronger magnetic fields at each spatial scale.
These visualizations therefore highlight the structure of the relatively strong magnetic field at each scale.
Both configurations, where the magnetic field lines are perpendicular and parallel to the streams, are present.
The details of the relationship between the gas flow direction and the magnetic field orientation will be presented in a subsequent paper.

%%In the third row of Figure~\ref{fig:4}, magnetic field lines are shown in red, in addition to the outflows and infall streamlines.
%%Magnetic field lines are (red lines) are also plotted in panels (d)--(f) of Figure~\ref{fig:4}. 
%%The field lines are integrated from the high-density regions near the center, which trace the stronger magnetic fields at each spatial scale.
%%Thus, the visualizations highlight the structure of the relatively strong magnetic field in each scale. 
%%Both configurations where the magnetic field lines are perpendicular and parallel to the streams are present. 
%%The details of the relationship between the gas flow direction and the magnetic field orientation will be presented in a companion paper.
%%In Figure~\ref{fig:4}(d), the field lines are roughly aligned with their initial direction (the $z$-axis), which appears horizontally in the figure.
%%In Figure~\ref{fig:4}(h), the field lines near the center are twisted by rotational motion, resulting in a complex configuration.
%%This panel shows that the magnetic field is aligned with the infall streams in some regions and nearly perpendicular in others, indicating no simple or uniform relationship between the stream direction and the magnetic field orientation.
%%In Figure~\ref{fig:4}(i), the magnetic field structure becomes even more complex.
%%In addition to the twisting caused by disk rotation, the outflow also contributes to the deformation of the field lines.
%%To enable detailed comparison with observations, synthetic observations will be necessary, which are beyond the scope of this study.

\section{Discussion}
%%%%%%
% Fig. 5
%%%%%%
\begin{figure*}
\begin{center}
\includegraphics[width=1.0\columnwidth]{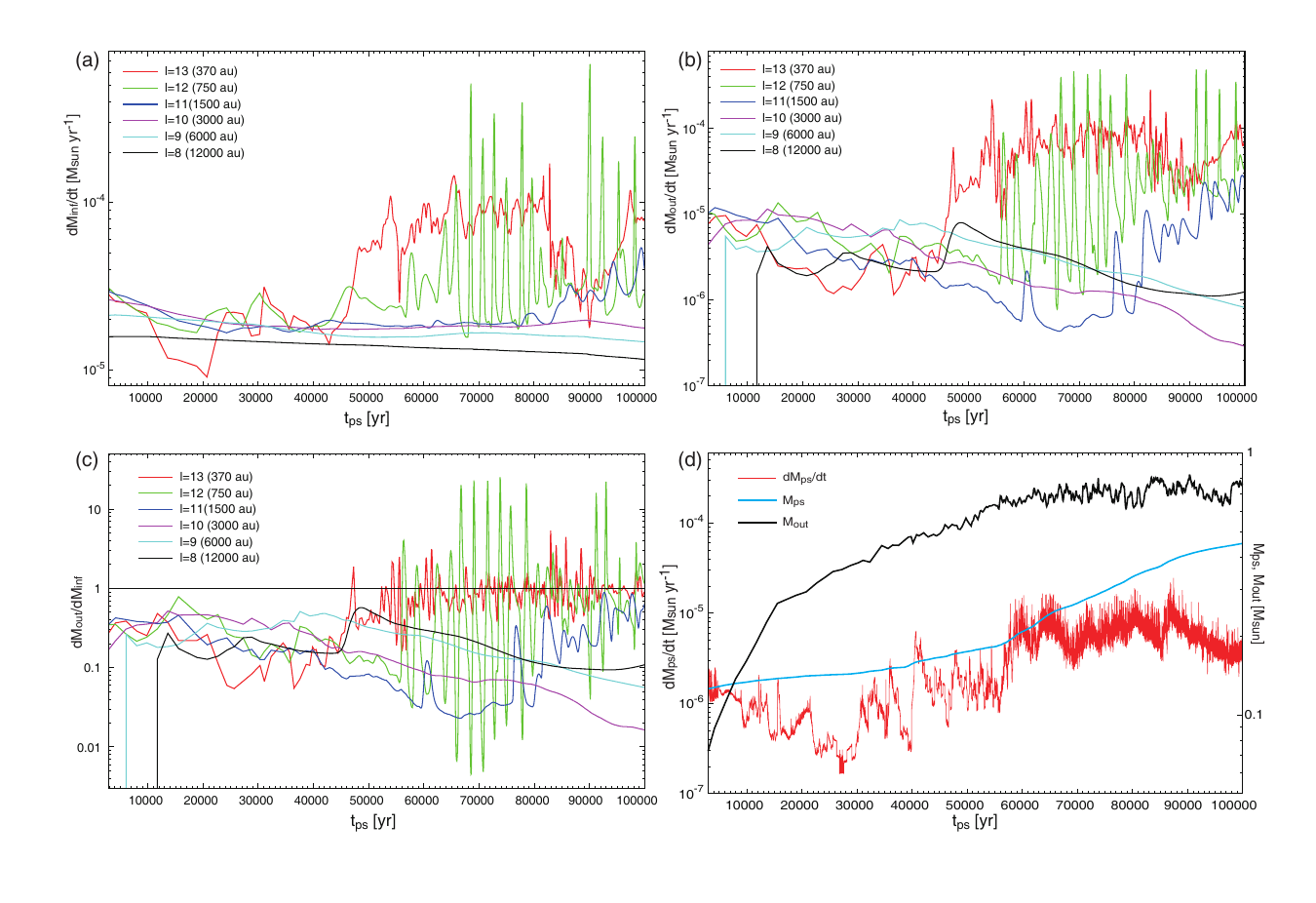}
\end{center}
\caption{
(a)–(c): Time evolution of the mass inflow rate $\dot{M}_{\rm inf}$ (panel (a)), the mass outflow rate $\dot{M}_{\rm out}$ (panel (b)), and the ratio of outflow to inflow rates $\dot{M}_{\rm out}/\dot{M}_{\rm inf}$ (panel (c)) at different spatial scales, corresponding to grid levels $l = 8$–13, plotted against the elapsed time $t_{\rm ps}$ after protostar formation.
(d): Time evolution of the mass accretion rate onto the protostar (sink) is shown on the left axis, while the protostellar mass and outflow mass are shown on the right axis, all as functions of $t_{\rm ps}$.
}
\label{fig:5}
\end{figure*}

In the classical picture, if strong initial turbulence is not assumed in the molecular cloud core, the mass accretion onto the disk and its surroundings is expected to remain nearly constant over short to moderate timescales, and then gradually decline over longer timescales \citep{Larson2003,vorobyov05b}.
In reality, however, the pseudodisk or infalling envelope undergoes continuous morphological changes.
This evolution is driven not only by magnetic and rotational effects but also by the development of protostellar outflows.
Therefore, even without considering external factors such as gas inflow or clump infall from outside the core, or internal factors such as initial turbulence, gas accretion onto the central region and the disk can be intrinsically time variable. 

Figure~\ref{fig:5}(a) shows the time variability of the mass infall (or accretion) rate at different spatial scales.
The mass infall  rate on each grid is estimated according to Equation (7) in \citet{machida24}.
In the regions near the disk (at 370\,au and 750\,au), the infall  rate onto the central region begins to fluctuate significantly after $t_{\rm ps} > 40,000$--$50,000$\,yr.
This variability arises from the twisted pseudodisk, which is created by magnetic field lines distorted by the rotation near the center.
In addition, magnetic braking facilitates inward gas motion by removing angular momentum, while the amplification and distortion of the magnetic field suppress accretion.
In regions where magnetic field lines are aligned straight toward the center, gas can accrete smoothly.
In contrast, when the field lines are twisted, the flow spirals along the distorted lines, resulting in a lower inflow  rate (or slower accretion) compared to the straight-line case.
As shown in Figure~\ref{fig:5}(a), the inflow rate at 1,500\,au begins to fluctuate after $t_{\rm ps} > 80,000$\,yr.
The evolution of the pseudodisk (or infalling envelope) and the magnetic field configuration proceeds from the inside out.
As a result, the time variability and spatial anisotropy of infall become more pronounced on larger scales over time.

Figure~\ref{fig:5}(b) shows the time variability of the mass outflow rate, and Figure~\ref{fig:5}(c) shows the ratio of the mass inflow rate to the mass outflow rate.
Similar to the mass inflow  rate, the mass outflow rate also exhibits strong time variability.
When the mass inflow  rate increases, a greater amount of gravitational energy is released by the infalling gas, which leads to an increase in the outflow rate \citep{matsushita17,Matsushita2018}. 
In addition, when a strong outflow appears, it can sweep up the infalling gas and eject it outward. 
As a result, the ratio of the mass outflow rate to the inflow  rate often exceeds unity, as seen in Figure~\ref{fig:5}(c).
Figure~\ref{fig:5}(c) also indicates that a nonnegligible fraction of the gas is ejected by the outflow.
%In our previous study, we showed that when the global magnetic field $B_0$ is aligned with the initial angular momentum vector $J_0$, the infall rate remains nearly constant even if the outflow is highly time-variable \citep{machida14,machida24}.
%However, as shown in this study, when $B_0$ and $J_0$ are misaligned, the outflow also appears to influence the mass accretion. 
The extent to which outflow affects the inflow  rate is still unclear, and we will explore this in a subsequent paper. 

\begin{figure*}
\begin{center}
\includegraphics[width=1.0\columnwidth]{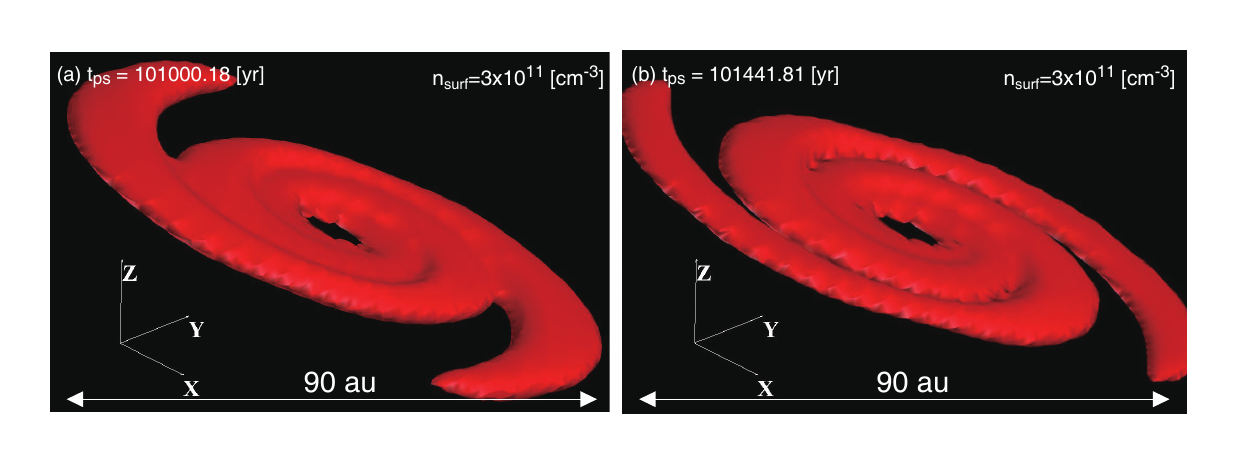}
\end{center}
\caption{
Three-dimensional view of the iso-density surface at a number density of $n_{\rm surf}=3.1\times10^{11}\,{\rm cm^{-3}}$ at $t_{\rm ps}=101{,}000.18\,{\rm yr}$ (a) and $t_{\rm ps}=101{,}441.81\,{\rm yr}$ (b). 
The high-density part of the disk is extracted to emphasize the nonaxisymmetric structure  around the protostar. 
The spatial scale is indicated in each panel.
}
\label{fig:6}
\end{figure*}

Figure~\ref{fig:5}(d) shows the mass accretion rate onto the protostar (sink) and the time evolution of the protostar and outflow masses. 
The accretion rate onto the central protostar is primarily controlled by the physical condition of the disk (such as magnetic field strength and Toomre's $Q$), but it is also affected by the accretion from the infalling envelope onto the disk. 
When the accretion onto the disk varies strongly over time, the accretion rate onto the protostar also shows significant time variability, as seen in Figures~\ref{fig:5}(a) and \ref{fig:5}(d).
However, the timescale of the mass inflow (or accretion) from the infalling envelope to the disk ($\gtrsim 1,000$\,yr) differs significantly from that of the accretion from the disk to the protostar ($\lesssim 100$\,yr). 
The short-term variability in the accretion rate onto the protostar is determined by the growth and decay of nonaxisymmetric structures caused by the disk gravitational instability \citep{machida10, Tomida17,machida19,hirano20,Machida2020d}  in the inner disk region.
Figure~\ref{fig:6} shows that, although the morphological change is not very large, the structure of the high-density part of the disk near the protostar changes  on a relatively short timescale of $\sim 100$\,yr. 
Therefore, the short-term variability ($\lesssim 100$\,yr) seen in Figure~\ref{fig:6}(d) should be attributed to the time-dependent structural evolution of the high-density inner disk, which strongly affects the mass accretion from the disk onto the protostar. 
On the other hand, variability on timescales of $\gtrsim 1,000$\,yr should be attributed to the mass infall from the twisted pseudodisk or complex infalling envelope. 
%%Therefore, changes in the accretion process on larger scales are expected to influence the evolution of both the disk and the protostar. 
Figure~\ref{fig:5}(d) also shows that the outflow mass exceeds the protostellar mass at all times, which indicates that a substantial amount of gas is ejected even when the infalling envelope has a complex structure.

%%%%%%
% Fig. 6
%%%%%%

\section{Summary}
We investigated the long-term evolution of an isolated, non-turbulent prestellar core with a 30$^\circ$ misalignment between the magnetic field and rotation axis.
The calculation followed the evolution for over 100,000\,yr after protostar formation.
A pseudodisk forms during collapse due to magnetic forces, enclosing the rotationally supported disk and acting as the infalling envelope. 
In the misaligned model, the magnetic field is strongly twisted after disk formation, producing a highly deformed pseudodisk.
This complexity spreads outward over time, and by 50,000\,yr, the pseudodisk shows irregular structures on scales of $\lesssim$ 5,000–10,000\,au. 
%%As a result, accretion and outflow patterns become strongly asymmetric and scale-dependent. 
%%On large scales, outflows align roughly with the initial magnetic field; on smaller scales ($\lesssim$3,000\,au), they follow the disk normal. 
%%In both regimes, high infall regions surround the outflow, indicating non-axisymmetric accretion.

The complex structure of the infalling envelope leads to substantial temporal variations in the mass inflow  rate onto the disk. 
%%These variations, in turn, influence both the mass accretion onto the central protostar and the evolution of the disk itself. 
Infall onto the disk occurs through multiple channels rather than uniformly.
We found that an asymmetric infalling envelope and asymmetric mass inflow  onto the disk naturally arise from the intrinsic dynamics of the collapsing core.
These asymmetric features appear even without external gas inflow, the infall of small clumps, or internal turbulence.
In other words, we demonstrate that asymmetric inflow can originate from internal processes.

\section*{Acknowledgements}
This research used the computational resources of the HPCI system provided by the Cyber Science Center at Tohoku University and the Cybermedia Center at Osaka University (Project ID: hp230035, hp240010, hp250007).
Simulations reported in this paper were also performed by 2024, 2025 Koubo Kadai on Earth Simulator (NEC SX-ACE) at JAMSTEC. 
The present study was supported by JSPS KAKENHI Grant (JP25K07369: MNM, JP21H01123 and JP21K13960: SH).
This work was supported by a NAOJ ALMA Scientific Research grant (No. 2022-22B). 
S.B. was supported by a Discovery Grant from NSERC.

\bibliography{machida}{}
\bibliographystyle{aasjournal}

\appendix
\section{Procedure to plotting streamlines}
\label{sec:A1}

%%%%%%
% Fig. A1
%%%%%%
\begin{figure*}
\begin{center}
\includegraphics[width=1.0\columnwidth]{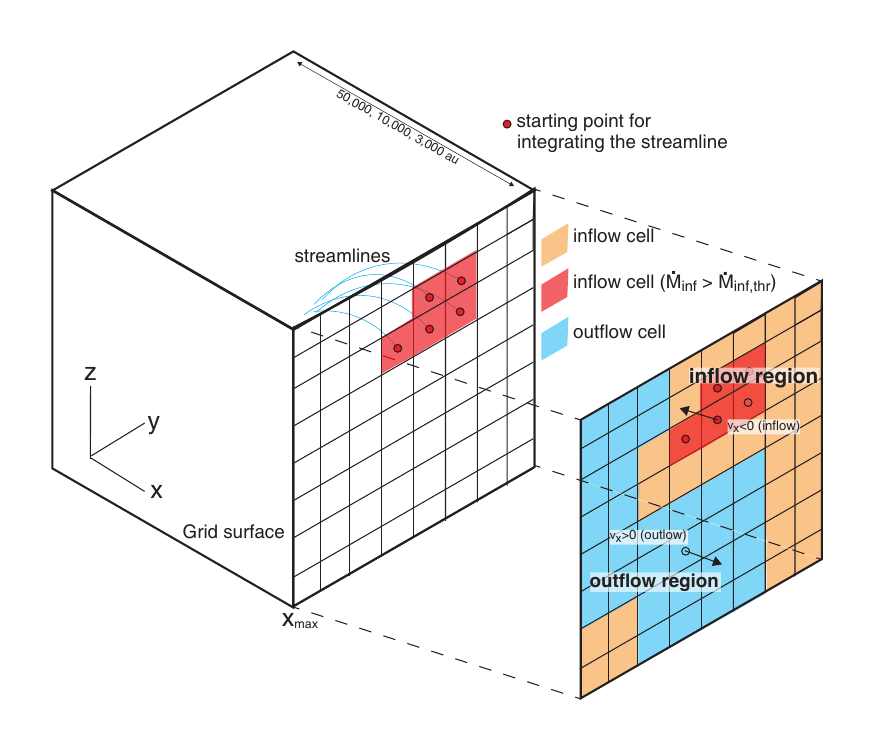}
\end{center}
\caption{
Schematic illustration of how the streamlines are plotted.
For clarity, the procedure is shown only on the grid surface of $x>0$ (or $x=x_{\rm max}$).
}
\label{fig:A1}
\end{figure*}

Figure~\ref{fig:A1} illustrates the procedure used to plot the streamlines in Figures~\ref{fig:4} and \ref{fig:A2}.
For clarity, the explanation is shown only for the surface at $x>0$ (or $x=x_{\rm max}$), but the same procedure is applied to all six outer surfaces of the Cartesian grid with 
box sizes of 50{,}000, 10{,}000, and 3{,}000 au.

As described in Section~\ref{sec:results}, inflow and outflow regions are 
identified on each surface using the velocity component normal to that surface 
(e.g., $v_x$ on the $x>0$ plane). Among the inflow cells, we select those whose 
mass inflow rate exceeds the threshold value $\dot{M}_{\rm inf,thr}$. 
Streamlines are then integrated inward from these selected surface cells using 
the full three-dimensional velocity vector at each grid cell. 
This procedure allows us to trace the dominant flow paths from the outer boundary toward the 
central region while avoiding the ambiguity associated with tangential fluxes between neighboring surface cells.

\section{Three-dimensional view of streamlines, magnetic field lines and outflows}
\label{sec:A2}
\begin{figure*}
\begin{center}
\includegraphics[width=1.0\columnwidth]{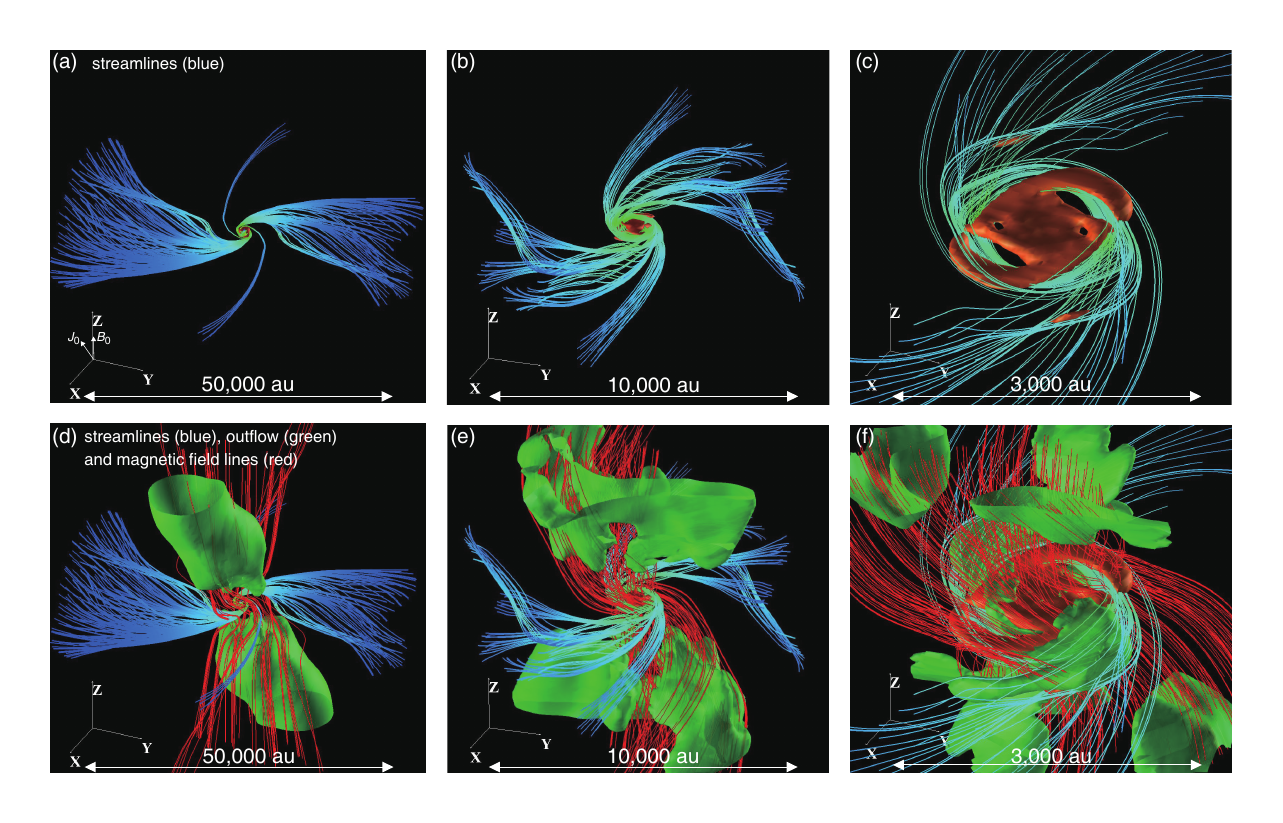}
\end{center}
\caption{
Same as Fig.~\ref{fig:4}, but shown from a different viewing angle. 
The initial directions of the magnetic field $B_0$ and angular momentum $J_0$ are indicated in panel (a).
}
\label{fig:A2}
\end{figure*}

Figure~\ref{fig:A2} is the same as Fig.~\ref{fig:4}, but viewed from different angles. 
For reference, the initial directions of the magnetic field $B_0$ and angular momentum $J_0$ are indicated in panel (a). 
Panels (d)--(f) show that the outflow (green surfaces) propagates roughly along the $z$-direction. 
The magnetic field lines (red) are also roughly aligned with the $z$-axis on large scales (panel d), while exhibiting a highly twisted and complex morphology on smaller scales (panel f). 
The streamlines (blue) indicate that the infalling gas approaches the central region mainly from lateral directions, approximately perpendicular to the initial $B_0$ (or equivalently, the $z$) direction.

%%%%%%
% Fig. A2
%%%%%%

\end{document}